\documentclass[superscriptaddress,preprintnumbers,amsmath,amssymb,twocolumn,floats]{revtex4-2}
\usepackage{txfonts}
\usepackage{amssymb}
\usepackage{amsfonts}
\usepackage{amsmath}
\usepackage{graphicx}
\usepackage{epstopdf}
\usepackage{color}
\begin{document}

\title{Large anomalous Hall effect in the kagome ferromagnet LiMn$_6$Sn$_6$}

\begin{abstract}
Kagome magnets are believed to have numerous exotic physical properties due to the possible interplay between lattice geometry, electron correlation and band topology. Here, we report the large anomalous Hall effect in the kagome ferromagnet LiMn$_6$Sn$_6$, which has a Curie temperature of 382 K and easy plane along with the kagome lattice. At low temperatures, unsaturated positive magnetoresistance and opposite signs of ordinary Hall coefficient for $\rho_{xz}$ and $\rho_{yx}$ indicate the coexistence of electrons and holes in the system. A large intrinsic anomalous Hall conductivity of 380 $\Omega^{-1}$ cm$^{-1}$, or 0.44 $e^2/h$ per Mn layer, is observed in $\sigma_{xy}^A$. This value is significantly larger than those in other $R$Mn$_6$Sn$_6$ ($R$ = rare earth elements) kagome compounds. Band structure calculations show several band crossings, including a spin-polarized Dirac point at the K point, close to the Fermi energy. The calculated intrinsic Hall conductivity agrees well with the experimental value, and shows a maximum peak near the Fermi energy. We attribute the large anomalous Hall effect in LiMn$_6$Sn$_6$ to the band crossings closely located near the Fermi energy.

\end{abstract}

\author{Dong Chen}
\email{Dong.Chen@cpfs.mpg.de}
\affiliation{Max Planck Institute for Chemical Physics of Solids, 01187 Dresden, Germany}
\affiliation{College of Physics, Qingdao University, Qingdao 266071, China}

\author{Congcong Le}
\affiliation{Max Planck Institute for Chemical Physics of Solids, 01187 Dresden, Germany}

\author{Chenguang Fu}
\affiliation{Max Planck Institute for Chemical Physics of Solids, 01187 Dresden, Germany}
\affiliation{State Key Laboratory of Silicon Materials, and School of Materials Science and Engineering, Zhejiang University, Hangzhou 310027, China}

\author{Haicheng Lin}
\affiliation{Max Planck Institute for Chemical Physics of Solids, 01187 Dresden, Germany}

\author{Walter Schnelle}
\affiliation{Max Planck Institute for Chemical Physics of Solids, 01187 Dresden, Germany}

\author{Yan Sun}
\affiliation{Max Planck Institute for Chemical Physics of Solids, 01187 Dresden, Germany}

\author{Claudia Felser}
\affiliation{Max Planck Institute for Chemical Physics of Solids, 01187 Dresden, Germany}

\maketitle

\section{Introduction}
The magnet with kagome lattice has been studied for a long time in the condensed matter physics and material science communities \cite{b1,b2,b3}. Kagome lattice is a two-dimensional honeycomb network of corner-sharing triangles. Owing to this special lattice geometry, a band structure with Dirac cones and a flat band has been revealed by the simple tight-binding model \cite{b4}. Embodied with magnetism and spin-orbit coupling (SOC), some kagome magnets have been recently found to have exotic topological states and phenomena. For example, the spin-polarized Dirac cone with a spin-orbit-coupling induced gap has been observed in kagome ferromagnet Fe$_3$Sn$_2$ \cite{b5}. Giant intrinsic anomalous Hall effect caused by the large Berry curvature have been observed in antiferromagnets Mn$_3$Sn, Mn$_3$Ge \cite{b6,b7,b8}, and the magnetic Weyl semimetal Co$_3$Sn$_2$S$_2$ \cite{b9,b10}. These experimental breakthroughs indicate that kagome magnet is a promising system for investigating the interplay of lattice geometry, electron correlation, and band topology.

Recently, another kagome magnet family $R$Mn$_6$Sn$_6$ ($R$ = trivalent rare earth elements) also has received much attention \cite{b11,b12,b13,b14}. They have the same layered hexagonal structure (space group P6/mmm), which is composed by kagome Mn$_3$ layers alternatively inserted by $R$Sn$_2$ and Sn$_4$ layers. All the members are magnetic ordered with relatively high transition temperatures \cite{b15,b16,b17,b18}. Among these materials, TbMn$_6$Sn$_6$, with a ferrimagnetic order perpendicular to the kagome lattice, was found to have the Chern gapped Dirac fermions \cite{b11}. Antiferromagnetic YMn$_6$Sn$_6$ was found to exhibit giant topological Hall effect near room temperature \cite{b12}. Although with multiple magnetic structures and fascinating properties, ferromagnetism is absent in $R$Mn$_6$Sn$_6$ family and the discovered anomalous Hall effect among them is not impressive. Actually, the $R$ elements can be completely replaced by Li, Mg, or Ca, and the magnetism of the new systems are all ferromagnetic \cite{b19,b20}. In the meantime, the lower valences of Li, Mg, and Ca also reduce the system's number of valence electrons. Thus, the simultaneously tuned magnetic and electronic states in these new ``166" compounds give us a chance to pursue larger anomalous Hall effect and other new phenomena. 

In this paper, we report a study on the magnetotransport properties of LiMn$_6$Sn$_6$, a member of the ``166" material family with minimal number of valence electrons. This compound is a ferromagnet with the Curie temperature ($T\rm_C$) of 382 K and has an easy plane parallel to the kagome lattice ($ab$ plane). The resistivity shows a metallic temperature dependence, and the low-temperature magnetoresistance is positive and has no tendency to saturate below 9 T. Hall resistivity shows opposite signs in the ordinary Hall coefficient of $\rho_{xz}$ and $\rho_{yx}$, indicating the coexistence of electrons and holes in the system. A large intrinsic contribution in the anomalous Hall conductivity (AHC) of 380 $\Omega^{-1}$ cm$^{-1}$, or 0.44 $e^2/h$ per Mn layer, can be obtained, which is much larger than those in the $R$Mn$_6$Sn$_6$ compounds. By band structure calculations, we find several linear band crossings and a small density of states (DOS) near the $E\rm_F$. At the K point of the Brillouin zone, we find a spin-polarized Dirac dispersion close to the $E\rm_F$. The energy dependent intrinsic AHC calculated from Berry curvature has a maximum peak near the $E\rm_F$ with the peak value close to the experimental one. These results show that the large anomalous conductivity in LiMn$_6$Sn$_6$ is related to the band crossings near the $E\rm_F$.

\section{Experiment and Methods}
The single crystals of LiMn$_6$Sn$_6$ were grown by self-flux method. Li, Mn, and Sn with atomic ratio of Li : Mn : Sn = 3 : 3 : 10 were loaded in a Ta crucible, and a Ta filter with drilled holes was then knocked into the crucible. The crucible was finally arc welded with a Ta cap and sealed in an evacuated quartz tube. The tube was heated to 1000 $^{\circ}$C, kept for 20 hours, and then cooled down to 500 $^{\circ}$C with a rate of 3 $^{\circ}$C/h. After that, the tube was centrifuged to separate the crystals from flux. Hexagonal plate-shaped single crystals were obtained without significant air sensitivity. The crystal structure of the samples was checked by x-ray diffraction (XRD) on a PANalytical diffractometer with Cu $K\alpha$ radiation at room temperature. The hexagonal plate-shaped crystals were used for the magnetic properties measurements, and were cut into sticks before the electrical transport measurements. The magnetic properties and electrical transport properties were measured on a Quantum Design Magnetic Properties Measurement System (MPMS) and a Physical Properties Measurement System (PPMS), respectively. 

\begin{figure}
	\includegraphics[width=8.6cm]{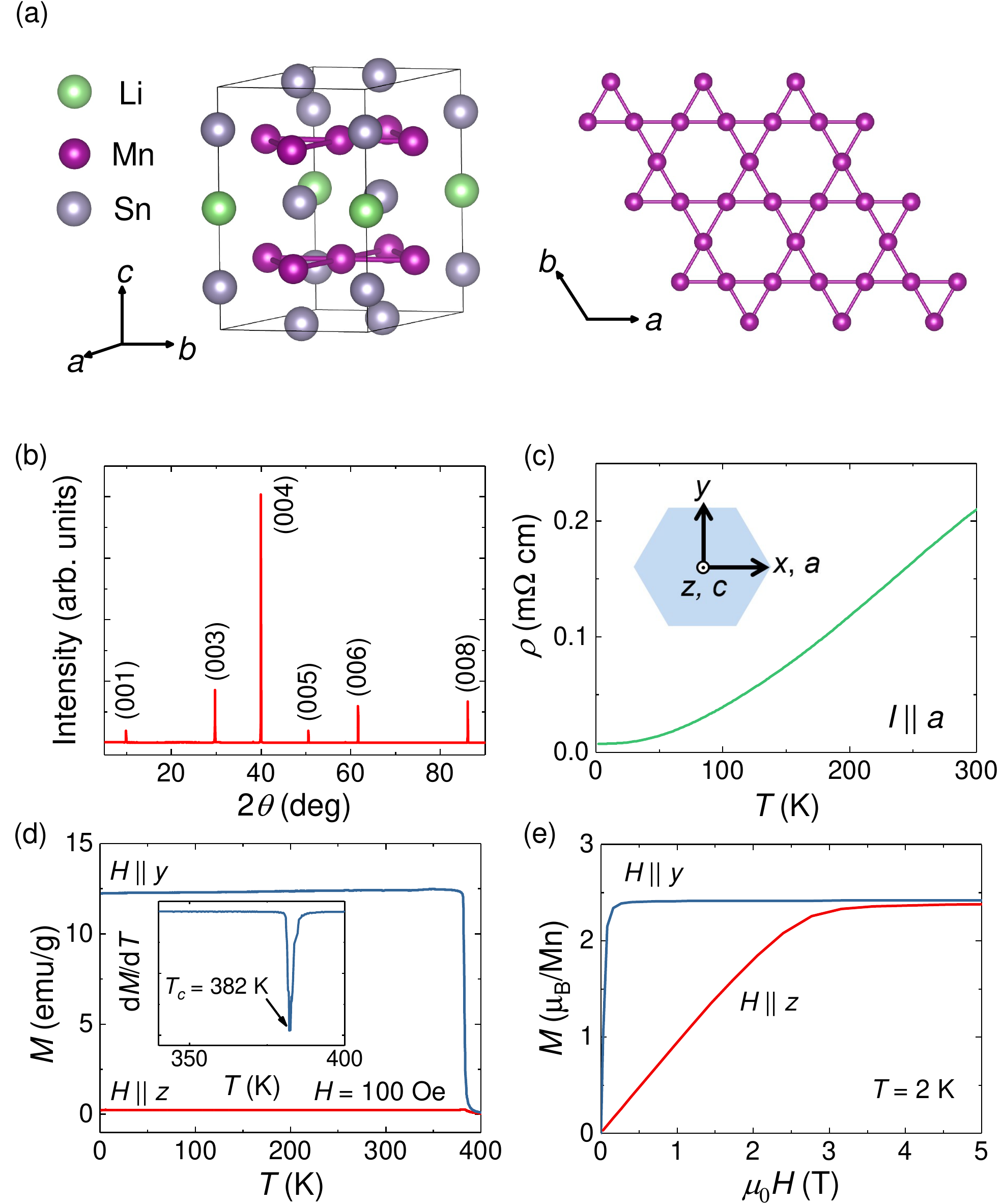}
	\caption{
		\label{f1}
		(a) Crystal structure of LiMn$_6$Sn$_6$. 
		(b) XRD pattern of a LiMn$_6$Sn$_6$ single crystal with (00$l$) reflections.  
		(c) Temperature dependence of resistivity with current along $a$ axis. The inset illustrates the definition of a Cartesian coordinate system based on the hexagonal lattice. 
		(d) Temperature dependence of the magnetization under magnetic field of $H$ = 100 Oe lying in and perpendicular to the $ab$ plane. The inset shows the derivative of the magnetization d$M$/d$T$ vs. $T$ for $H \parallel y$.
		(e) The isothermal magnetization at $T$ = 2 K for $H \parallel y$ and $H \parallel z$.
	}
\end{figure}

Our calculations were performed using density functional theory (DFT) as implemented in the Vienna ab initio simulation package (VASP) code \cite{b21,b22,b23}. The generalized-gradient approximation (GGA) for the exchange correlation functional was used. Throughout the work, the cutoff energy was set to be 550 eV for expanding the wave functions into plane-wave basis. In the calculation, the Brillouin zone was sampled in the $k$ space within Monkhorst-Pack scheme \cite{b24}. On the basis of the equilibrium structure, the $k$ mesh used was $10\times10\times6$.  The intrinsic Hall conductivity $\sigma_{xy}^{int}$ is obtained by integrating the $z$ component of Berry curvature $\Omega_{xy}^z(\textbf{k})$ on all the occupied states through the Brillouin zone with SOC included and the moment of Mn along $z$ axis.

\section{Results and Discussion}

LiMn$_6$Sn$_6$ has the same crystal structure with the $R$Mn$_6$Sn$_6$ compounds, with lattice parameters of $a = b$ = 5.497 $\AA$, $c$ = 9.026 $\AA$. As shown in Fig. \ref{f1}(a), the Mn kagome lattice is parallel to the $ab$ plane. Figure \ref{f1}(b) shows the XRD pattern of the hexagonal surface of a plate-like single crystal. All the peaks can be identified as the (00$l$) reflections of LiMn$_6$Sn$_6$ as labeled on the pattern, indicating the (001) surface of the crystal. The phase of the samples was further checked by the powder XRD. Figure \ref{f1}(c) shows the temperature dependence of the resistivity with current flowing along $a$ axis. It has a metallic behavior with a residual resistivity ratio $\rho$(300 K)/$\rho$(2 K) = 29. The inset of Fig. \ref{f1}(c) defines a Cartesian coordinate system for later using, with $x \parallel a$, $z \parallel c$, and $y \perp x, z$. The ferromagnetism of the sample can be illustrated by the temperature dependence of magnetization with magnetic field $H$ = 100 Oe along with the $ab$  plane ($H \parallel y$) and the $c$ axis ($H \parallel z$) [Fig. \ref{f1}(d)]. The magnetization for $H \parallel y$ is much larger than that for $H \parallel z$ below $T_C$, suggesting the $ab$ plane is the easy plane. The $T_C$ can be determined by the peak of the derivative of magnetization d$M$/d$T$ for $H \parallel y$, as shown in the inset of Fig. \ref{f1}(d), which gives $T_C$ = 382 K. Figure \ref{f1}(e) shows the isothermal magnetization curves at $T$ = 2 K for the two directions. Due to the easy-plane anisotropy and the high quality of the crystals, there is no hysteresis in the magnetization curve. The saturation magnetization is about 2.4 $\mu_B$/Mn. All the magnetic properties are consistent with those previously revealed by neutron scattering experiment \cite{b19}.

\begin{figure}
	\includegraphics[width=8.6cm]{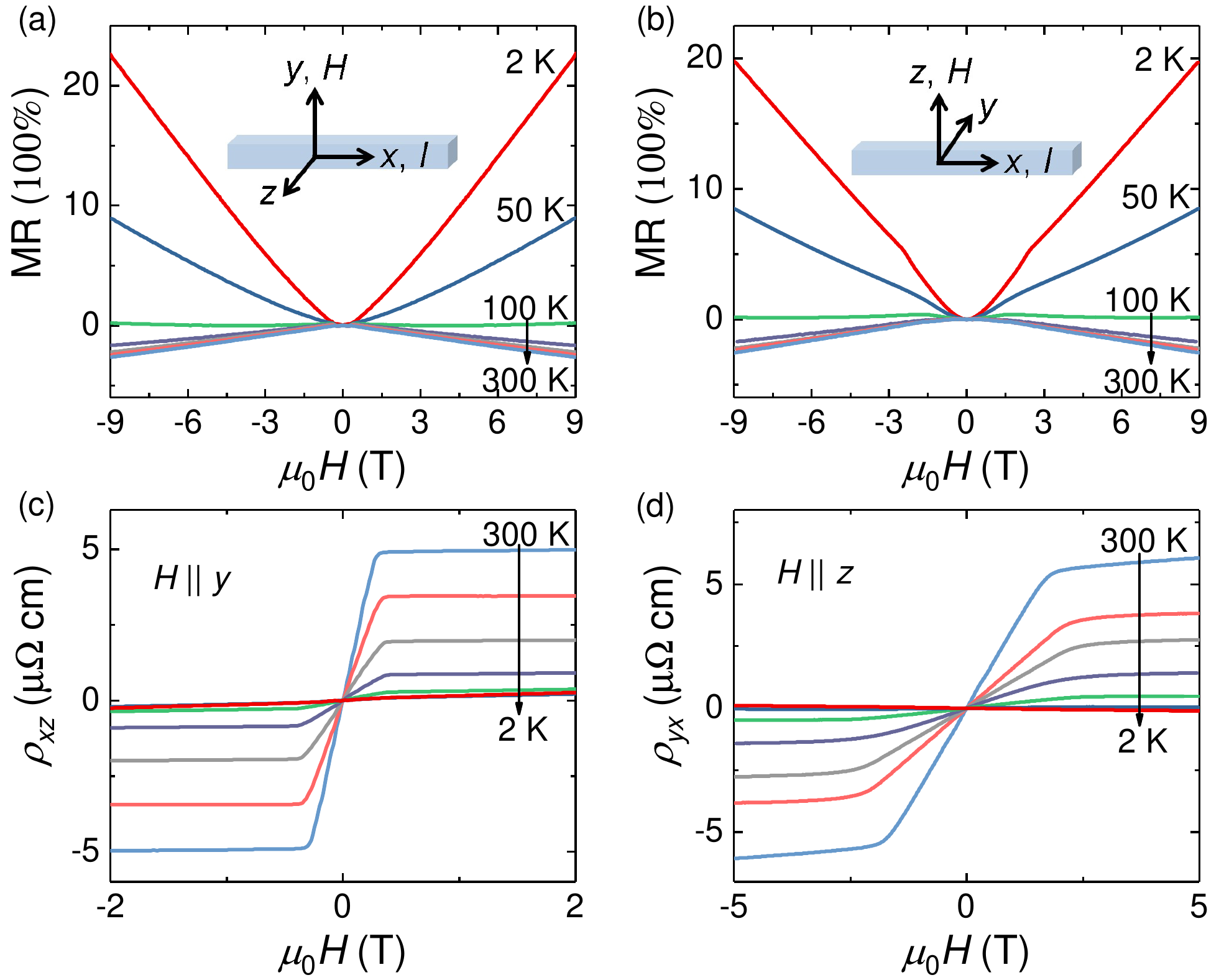}
	\caption
	{
		\label{f2}
		(a, b) Magnetic field dependence of magnetoresistance at various temperatures measured with $H \parallel y$ and $H \parallel z$, respectively. The insets show the corresponding measurement geometries. 
		(c, d) Magnetic field dependence of Hall resistivity at various temperatures with the same measurement geometries with (a) and (b), respectively.
	}
\end{figure}

Figures \ref{f2}(a) and (b) show the magnetic field dependence of the MR with the measurement geometries illustrated in the corresponding insets. The current is always along $x$ axis, and the magnetic field is along $y$ and $z$ axis, respectively. The MR for $H \parallel y$ shows a parabolic-like field dependence at $T$ = 2 K, and has no tendency to saturate below 9 T.  For $H \parallel z$, the MR at low temperatures show a kink around 2 T, and has a nearly linear behavior in higher field. With temperature increases, the MR for both directions gradually decline and become negative when temperature higher than 100 K. As a ferromagnetic metal, the negative MR of LiMn$_6$Sn$_6$ at high temperature can be understood as the consequence of suppressed spin-disorder scattering. While at low temperature, the positive MR indicates that some other mechanisms dominate in the transport process. 

To extract more information in the transport process, we performed the Hall effect measurements. Figures \ref{f2}(c) and (d) show the field dependence of the Hall resistivity $\rho_{xz}$ and $\rho_{yx}$ for $H \parallel y$ and $H \parallel z$, respectively, which are defined as $\rho_{xz} = -E_z/j_x$ and $\rho_{yx} = E_y/j_x$. At high temperatures, the Hall resistivity in both directions exhibit obvious anomalous Hall effect. Conventionally, the Hall resistivity of ferromagnets can be expressed as $\rho_{H} = \rho_{H}^0 + \rho_{H}^A = R_0B + R_s\mu_0M$, where $\rho_{H}^0$ ($\rho_{H}^A$) and $R_0$ ($R_s$) are the ordinary (anomalous) Hall resistivity and coefficient, respectively. It can be seen that the anomalous Hall resistivity has different saturation fields with the magnetization curves shown in Fig. \ref{f1}(e), especially for $H \parallel y$. This is caused by the different demagnetization factors that come from the different sample shapes in the two measurements. The ordinary Hall resistivity shows different slopes in the two directions. This is more obvious at lower temperatures. At 2 K, the anomalous Hall resistivity becomes too weak to be observed due to the reduction of resistivity, and the ordinary Hall resistivity is dominant. At this temperature, the slope of Hall resistivity is positive for $\rho_{xz}$ while negative for $\rho_{yx}$, which means that holes (electrons) are dominant for $H \parallel y$ ($H \parallel z$). This phenomenon reflects the coexistence of both holes and electrons in this compound, and the densities can be tuned by the direction of magnetic field. The coexistence of holes and electrons may explain the positive unsaturated MR at low temperature, which may be caused by the compensation effect of the two types of carrier \cite{b25}.

\begin{figure}
	\includegraphics[width=8.6cm]{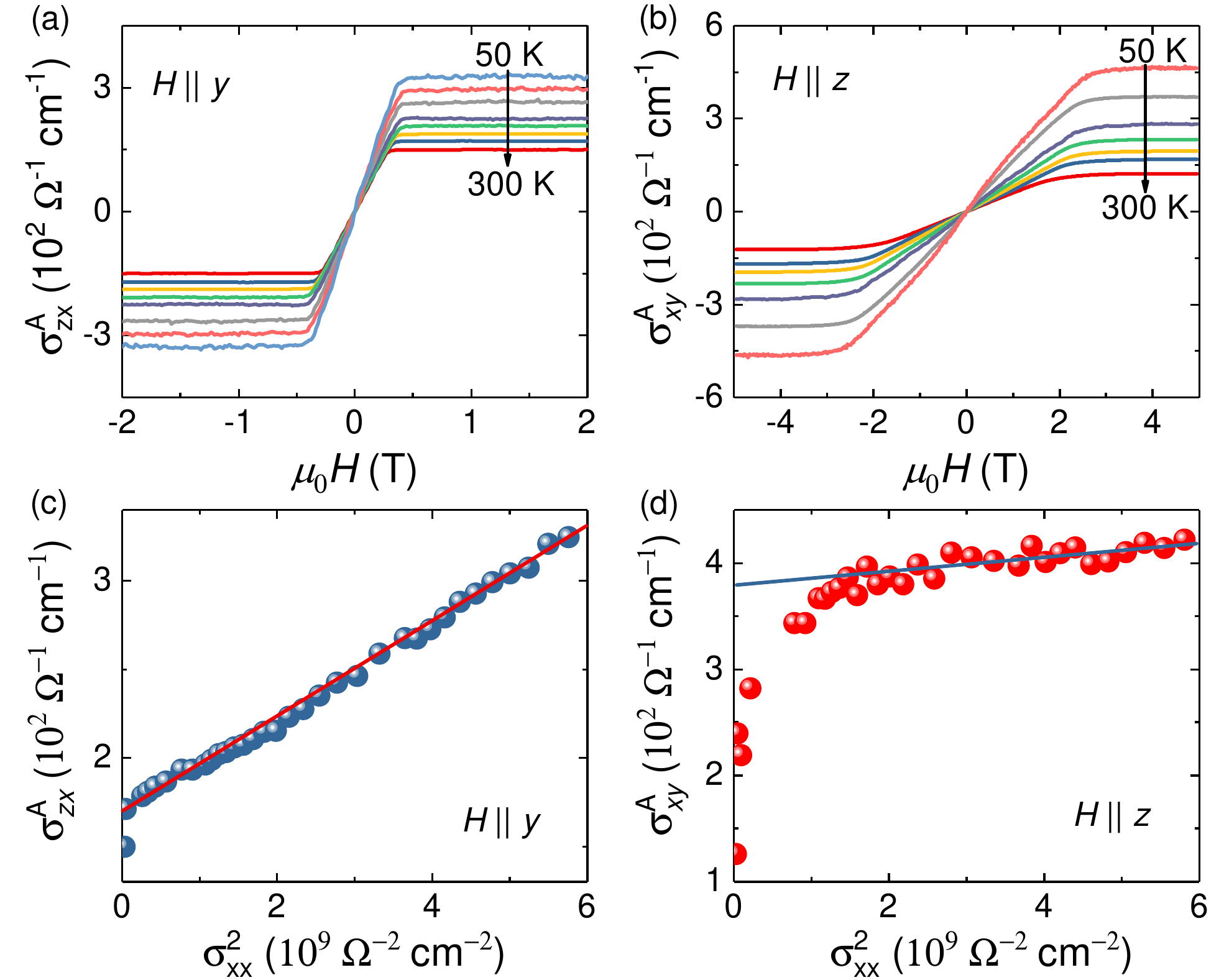}
	\caption
	{
		\label{f3}
		(a, b) Anomalous part of the Hall conductivity $\sigma_{zx}^A$ and $\sigma_{xy}^A$ as the function of magnetic field at various temperatures measured with $H \parallel y$ and $H \parallel z$, respectively. 
		(c, d) The anomalous Hall conductivity $\sigma_{zx}^A$ and $\sigma_{xy}^A$ versus $\sigma_{xx}^2$ with fitting lines, respectively.
	}
\end{figure}

The anomalous Hall effect in ferromagnetic conductor is considered to come from three different contributions, and the AHC can be expressed as $\sigma_{H}^A = \sigma_{int} + \sigma_{sk} + \sigma_{sj}$, where $\sigma_{int}$ is the intrinsic contribution, and $\sigma_{sk}$ and $\sigma_{sj}$ are the parts come from screw scattering and side jump mechanisms, respectively \cite{b26}. Both of the $\sigma_{sk}$ and $\sigma_{sj}$ terms are the consequences of extrinsic scattering from disorder or impurities, while the $\sigma_{int}$ term depends only on the band structure. Figures \ref{f3}(a) and (b) show the anomalous parts of the Hall conductivity $\sigma_{zx}$ and $\sigma_{xy}$, respectively, which are obtained by $\sigma_{ij} = \rho_{ji}/(\rho_{ji}^2+\rho_{ii}^2)$. To extract the intrinsic contribution of the AHC, we employ the so-called Tian-Ye-Jin (TYJ) scaling \cite{b5,b27}: $\sigma_{xy}^A = a\sigma_{xx}^2+\sigma_{xy}^{int}$, where $a$ is a system relevant constant, and $\sigma_{xy}^{int}$ is the intrinsic AHC contribution. Figures \ref{f3}(c) and (d) show the AHC as the function of $\sigma_{xx}^2$ for the two directions. The intrinsic contribution of AHC can be obtained from the intercept of the fitting line, which is 170 $\Omega^{-1}$ cm$^{-1}$ and 380 $\Omega^{-1}$ cm$^{-1}$ for $\sigma_{zx}^{int}$ and $\sigma_{xy}^{int}$, respectively. This anisotropic intrinsic AHC suggests the influence of magnetic field direction on the electronic structure of LiMn$_6$Sn$_6$ again. The value of $\sigma_{xy}^{int}$ can be converted into 0.88 $e^2/hc$, where $c$ is the cross-plane lattice constant. Considering there are two kagome Mn layers in a unit cell, it gives 0.44 $e^2/h$ per Mn layer. This intrinsic Hall conductivity component in LiMn$_6$Sn$_6$ is much larger than those that have been found in the $R$Mn$_6$Sn$_6$ compounds with ferrimagnetic order, such as 0.27 $e^2/h$ per Mn layer for GdMn$_6$Sn$_6$, and 0.14 $e^2/h$ per Mn layer for TbMn$_6$Sn$_6$, as listed in Table \ref{t1}.

\begin{table}
	\caption{Comparison of the intrinsic AHC of LiMn$_6$Sn$_6$ with those of $R$Mn$_6$Sn$_6$, where FIM denotes the ferrimagnet and FM is the ferromagnet.}
	\label{t1}
	\setlength{\tabcolsep}{2mm}
	\begin{tabular}{lccc}
		\hline\hline
		&Magnetic order&$\sigma_{xy}^{int}$ ($e^2/h$ per Mn layer)&Reference\\
		\hline
		Gd&FIM&0.27, 0.17&\cite{b13,b14}\\
		Tb&FIM&0.14&\cite{b11,b14}\\
		Dy&FIM&0.07&\cite{b14}\\
		Ho&FIM&0.13&\cite{b14}\\
		Er&FIM&0.04&\cite{b14}\\
		Li&FM&0.44&This work\\
		\hline\hline
	\end{tabular}
\end{table}

\begin{figure}
	\includegraphics[width=8.6cm]{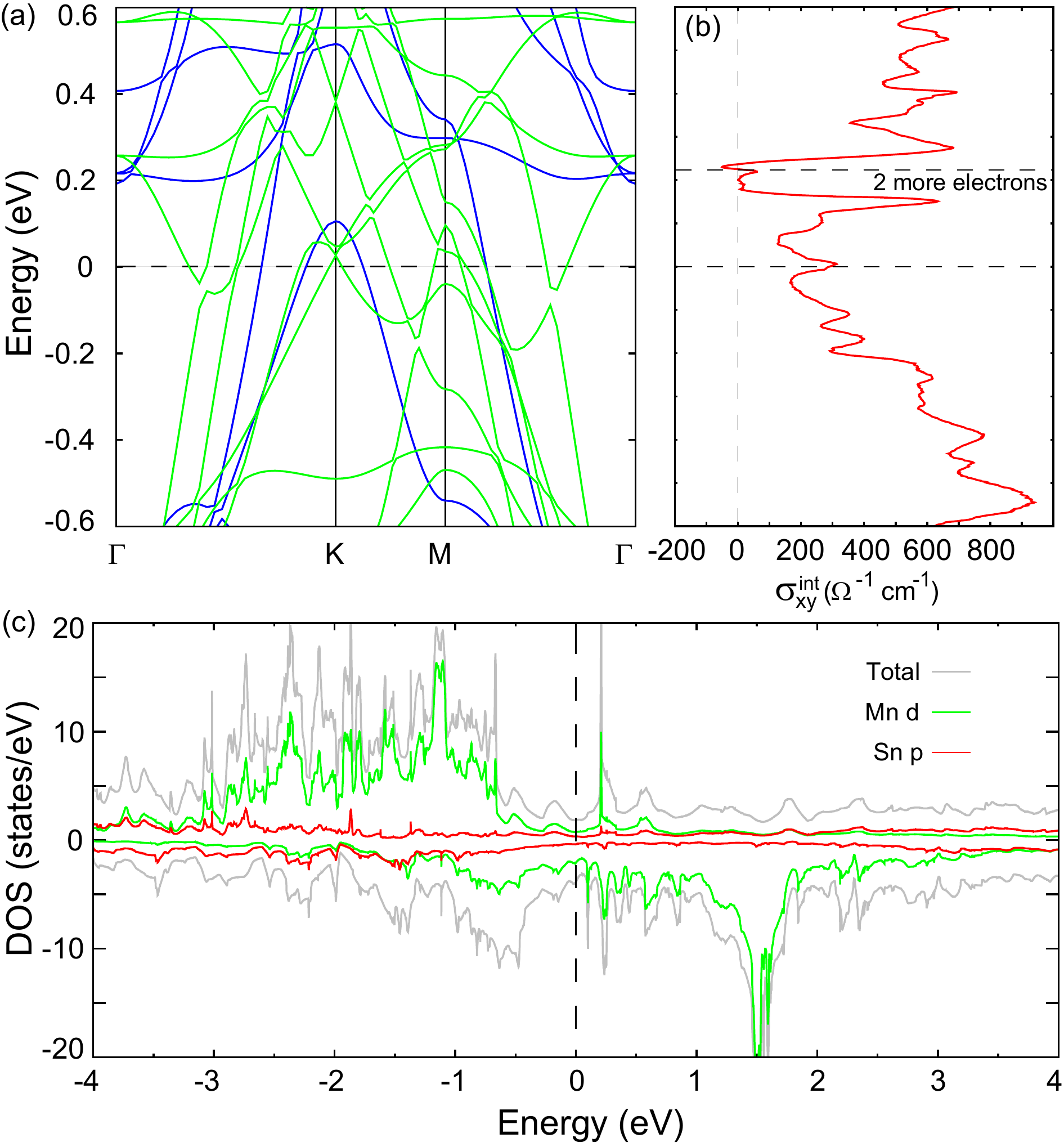}
	\caption
	{
		\label{f4}
		(a) The band structure of LiMn$_6$Sn$_6$ without SOC. The spin-up bands appear in blue lines, and the spin-down bands appear in green lines.
		(b) Energy dependence of the intrinsic AHC $\sigma_{xy}^{int}$, with the $E\rm_F$ of LiMn$_6$Sn$_6$ is 0. The upper dashed line denotes the lifted $E\rm_F$ with extra 2 electrons per formula unit.
		(c) Spin-resolved total and orbital-projected densities of states in the vicinity of $E\rm_F$.
	}
\end{figure}

Because $\sigma_{int}$ depends only on the band structure, we carried out the first-principles calculations to have a better understanding of the large $\sigma_{xy}^{int}$ in LiMn$_6$Sn$_6$. Figure \ref{f4} (a) shows the band structure without SOC in the vicinity of $E\rm_F$. Both spin-up and spin-down bands pass through the $E\rm_F$, with several crossings close to the $E\rm_F$. If the SOC is included into account, the band structure is almost unaltered due to the light elements, only with some small gaps opened in the band crossings. This band structure is consistent with the two types of carriers revealed by the Hall measurements. It is worth pointing out that there is a spin-polarized linear crossing slightly above the $E\rm_F$ at the K point, and it is gapped with the consideration of SOC. This is reminiscent of the Chern gapped Dirac fermions found in TbMn$_6$Sn$_6$ \cite{b11}. Figure \ref{f4}(c) shows the spin-resolved density of states near the $E\rm_F$. The bands near the $E\rm_F$ are mainly hybridized by the Mn-$d$ and Sn-$p$ orbitals. The DOS for both directions have relatively small values at the $E\rm_F$, consistent with the several band crossings near the $E\rm_F$ revealed by the band structure. The calculated magnetic moment based on the spin-resolved DOS is 2.4 $\mu_B$/Mn, same with the value obtained in the magnetization curves.  The energy dependence of intrinsic AHC $\sigma_{xy}^{int}$ can be calculated from Berry curvature based on the band calculations. As shown in Fig. \ref{f4} (b), there is a peak near the $E\rm_F$ with the value at $E\rm_F$ of 300 $\Omega^{-1}$ cm$^{-1}$, which is close to the experimental value. In the time-reversal-symmetry-broken systems, a band crossing near the $E\rm_F$ with a SOC induced gap is believed to contribute a large Berry curvature \cite{b28}. Thus, the peak of $\sigma_{xy}^{int}$ near $E\rm_F$ can be understood as the consequence of the close positions of the band crossings to the $E\rm_F$. We can also notice a higher peak of $\sigma_{xy}^{int}$ centered at about 0.15 eV, which is followed by a deep valley around 0.2 eV. If we assume the $R$Mn$_6$Sn$_6$ compounds have similar band structure, the $E\rm_F$ can be lifted for about 0.23 eV by the 2 extra valence electrons of $R$. This results in the smaller $\sigma_{xy}^{int}$ in $R$Mn$_6$Sn$_6$, as shown by the upper dashed line in Fig. \ref{f4}(b). All the calculation results fit well with the experimental observations, suggesting the reliability of our calculations. Based on the calculations, a much larger AHC up to 0.73 $e^2/h$ per Mn layer can be expected if the ferromagnetic compounds with $E\rm_F$ located at about 0.15 eV can be synthesized, such as (Li, $R$)Mn$_6$Sn$_6$ and (Mg, $R$)Mn$_6$Sn$_6$.

\section{Conclusion}
In conclusion, we have investigated the magnetic, transport properties, and band structure of LiMn$_6$Sn$_6$, which is a ferromagnet with $T_C$ = 382 K and the easy plan along its kagome lattice. With $H \parallel y$ and $H \parallel z$, LiMn$_6$Sn$_6$ shows nonsaturated positive MR and opposite ordinary Hall coefficients at low temperatures, suggesting the coexistence of electrons and holes. An intrinsic Hall conductivity component of 380 $\Omega^{-1}$ cm$^{-1}$ can be found in $\sigma_{xy}^{A}$, which is much larger than those in the ferrimagnetic $R$Mn$_6$Sn$_6$ compounds. First principles calculations reveal several band crossings including a spin-polarized Dirac dispersion, a minimum DOS, and a maximum intrinsic AHC near the $E\rm_F$. All of these results suggest the relationship between the large intrinsic anomalous Hall effect with the band crossings near the $E\rm_F$.

\section{Acknowledgments}
This work was supported by the European Research Council Advanced Grant (No. 742068) ``TOPMAT", the European Union's Horizon 2020 Research and Innovation Programme (No. 824123) ``SKYTOP", the European Union's Horizon 2020 Research and Innovation Programme (No. 766566) ``ASPIN", the Deutsche Forschungsgemeinschaft (Project-ID No. 258499086) ``SFB 1143", the Deutsche Forschungsgemeinschaft (Project-ID No. FE 633/30-1) ``SPP Skyrmions", and the DFG through the Wurzburg-Dresden Cluster of Excellence on Complexity and Topology in Quantum Matter ct.qmat (EXC 2147, Project-ID No. 39085490).

\end{document}